# Convolution filter embedded quantum gate autoencoder

Kodai Shiba*[1,3], Katsuyoshi Sakamoto[1], Koichi Yamaguchi[1], Dinesh Bahadur Malla[2] and Tomah Sogabe*[1,2,3]

*[1]Department of Engineering Science, The University of Electro-Communications, Tokyo,Japan,

*[2] i-PERC, The University of Electro-Communications, Tokyo, Japan

*[3] Technology Solution Group, Grid Inc., Tokyo, 107-0061, Japan

**Abstract**

The autoencoder is one of machine learning algorithms used for feature extraction by dimension reduction of input data, denoising of images, and prior learning of neural networks. At the same time, autoencoders using quantum computers are also being developed. However, current quantum computers have a limited number of qubits, which makes it difficult to calculate big data. In this paper, as a solution to this problem, we propose a computation method that applies a convolution filter, which is one of the methods used in machine learning, to quantum computation. As a result of applying this method to a quantum autoencoder, we succeeded in denoising image data of several hundred qubits or more using only a few qubits under the autoencoding accuracy of 98%, and the effectiveness of this method was obtained. Meanwhile, we have verified the feature extraction function of the proposed autoencoder by dimensionality reduction. By projecting the MNIST data to two-dimension, we found the proposed method showed superior classification accuracy to the vanilla principle component analysis (PCA). We also verified the proposed method using IBM Q Melbourne and the actual machine failed to provide accurate results implying high error rate prevailing in the current NISQ quantum computer.

## 1. Introduction

The quantum gate type quantum computer can construct a quantum circuit by combining quantum gates according to a problem. Therefore, it is versatile and will be put to practical use in several decades. However, the qubits of the quantum gate quantum computer are very weak to external interference, and it is difficult to maintain the quantum state for a long time. Therefore, as the number of qubits increases, noise is likely to be included, and the error correction function cannot be implemented. Therefore, the quantum gate type quantum computer NISQ (Noisy Intermediate-Scale Quantum computer), which has no error correction function and has a limited number of qubits, is currently being developed [1].

In addition, in a quantum simulator which can reproduce the quantum calculation in the quantum gate type quantum computer by the conventional classical computer, the calculation time increases exponentially as the number of qubits increases. Thus, computing big data that requires the use of a large number of qubits takes considerable time.

In this paper, as a solution to this problem, we propose a calculation method that applies convolution, which is one of the methods used in machine learning, to quantum computation. As an example, this method is applied to a quantum autoencoder using quantum adiabatic algorithm, which is one of quantum algorithms. Then, we consider the effectiveness of performing quantum computation of several hundred qubits or more using a convolution filter composed of several qubits.

In addition, this research is calculated using ReNomQ (https://github.com/ReNom-dev-team/ReNomQ) of a quantum gate type quantum simulator.

## 2. Quantum autoencoder

An autoencoder is a supervised learning algorithm that learns that output data reproduce input data [2]. It is composed of a three-layer neural network of an input layer, a hidden layer, and an output layer, and the data of the hidden layer is a data representing the features of the input data.

The quantum autoencoder is not a neural network but is calculated using quantum adiabatic algorithm, which is one of optimization algorithms.

### 2.1 Quantum adiabatic algorithm

Quantum adiabatic algorithm is one of the annealing calculation methods calculated using Ising model [3][4]. The Ising model is a model of the behavior of spins in a magnetic substance such as a ferromagnetic substance or an antiferromagnet, and has two types of states: up spin ($s_i = +1$) and down spin ($s_i = -1$).

The Hamiltonian of the entire system of the Ising model is expressed by the following equation using a coupling coefficient $J_{ij}$ between two spins $s_i$ and $s_j$, and a local longitudinal magnetic field $h_{z_i}$ applied to the inside of the spin $s_i$.

$$H = \sum_{i<j} J_{ij} s_i s_j + \sum_i h_{z_i} s_i \tag{1}$$

In quantum adiabatic algorithm, a transverse magnetic field $h_x$ is added to set the initial state of the Hamiltonian. Further, the spin $s_i$ corresponds to the Pauli operator $\sigma_i^z$, so it can be represented by the phase reversal operation gate $Z$ of the quantum gate represented by the same matrix. Furthermore, a parameter $s(= t/t_f)$ in which time $t$ is normalized to $t_f$ is introduced, and the interval is $0 \leq s \leq 1$. Therefore, the Hamiltonian in adiabatic quantum computation is as follows.

$$H(s) = s\left[\sum_{i<j} W_{ij}(\theta) Z_i Z_j + \sum_i h_{z_i}(\theta) Z_i\right] + (1-s) \sum_i h_x(\theta) X_i \tag{2}$$

$J_{ij}, h_{z_i}$ and $h_x$ are phase parameters of the unitary rotary gate. Therefore, they will be expressed as $W_{ij}(\theta), h_{z_i}(\theta)$ and $h_x(\theta)$. In quantum computers, it is possible to represent the time evolution of the Schrodinger equation by performing unitary transformations in order. Assuming that the qubit state vector is $|\psi\rangle$, Schrodinger equation can be expressed as follows.

$$i\hbar \frac{\partial}{\partial t} |\psi\rangle = H |\psi\rangle \tag{3}$$

Solving the Schrodinger equation in case of the state vector depends on time and the Hamiltonian does not depend on time, the following transformation can be made, and a unitary transformation $U(t)$ is derived.

$$|\psi(t)\rangle = e^{-i\frac{H}{\hbar}t} |\psi(0)\rangle \tag{4}$$

$$U(t) = e^{-i\frac{H}{\hbar}t} \tag{5}$$

Therefore, by substituting the eq. (2) into H in the eq. (5) and repeating the unitary transformation $U(t)$, we can obtain the minimum value of the Hamiltonian and obtain an optimal spin state.

In addition, a part of the quantum circuit of the eq. (5) is as shown in fig. 1, which is based on the book of Shigeru Nakayama [4]. In addition, Qiskit (https://qiskit.org/) of a quantum simulator is used for drawing a circuit diagram.

### 2.2 Normal quantum autoencoder

First, in order to verify the performance of a normal quantum autoencoder, we'll verifiy whether the same data as the original data can be output when the 9pixel black-and-white image is used as the original data and the noise mixed data obtained by adding noise to the original data is used as the input data.

Corresponding authors: sogabe@uec.ac.jp , shiba.kodai@gridsolar.jp

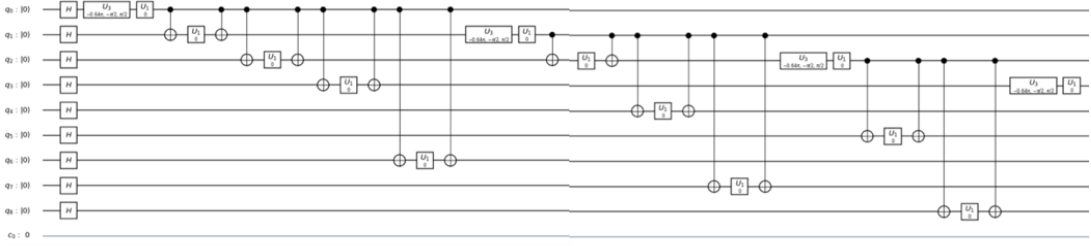

Fig. 1 A part of the quantum circuit diagram of a 9qubit quantum adiabatic algorithm in a quantum autoencoder

It is assumed that qubits correspond to each pixel, and qubits correspond to spins of Ising model. The characteristic data of the original data is assumed to be the coupling coefficient $W(\theta)$ and the longitudinal magnetic field coefficient $h_z(\theta)$, where $h_z(\theta)$ is directly set from the original data, and $W(\theta)$ is learned by the quantum autoencoder before noise removal. The setting of $h_z(\theta)$ is $h_{zi}(\theta) = -0.1$ when the $i$-th pixel of the original image is white, and $h_{zi}(\theta) = 0.1$ when black. Also, learning of $W(\theta)$ is performed by making the initial value of $W(\theta)$ random and updating after performing the quantum adiabatic algorithm. After performing the quantum adiabatic algorithm, $W(\theta)$ compares the output data with the original data and updates them so that they are equal. The update equation of the coupling coefficient $W_{ij}(\theta)$ between two spins $s_i$ and $s_j$ can be expressed as follows.

$$W_{ij}(\theta) = W_{ij}(\theta) - \varepsilon E_{ij} \quad (6)$$

$$E_{ij} = (s_i s_j)_{original} - s_i s_j \quad (7)$$

$E_{ij}$ is an error between the product of the two spins $s_i$ and $s_j$ of the original data and the product of the two spins $s_i$ and $s_j$ of the output data, and $\varepsilon$ is a learning rate. In this study, $\varepsilon$ was set to 0.4.

The spin, as shown in fig. 2(a), is considered only the coupling with the nearest four spins in the top, bottom, left, and right, and the periodic boundary condition is applied. Next, an image in which noise is mixed into the original image is prepared and used as input data for noise removal of the quantum autoencoder. The original image and the image after noise mixing are shown in fig. 2(b) and 2(c).

The result of having performed noise removal using a quantum adiabatic algorithm in this noise image is shown in fig. 2(d). In addition, fig. 2(e) shows the result of testing the noise removal 100 times and verifying the accuracy of the quantum autoencoder of the 9pixel image.

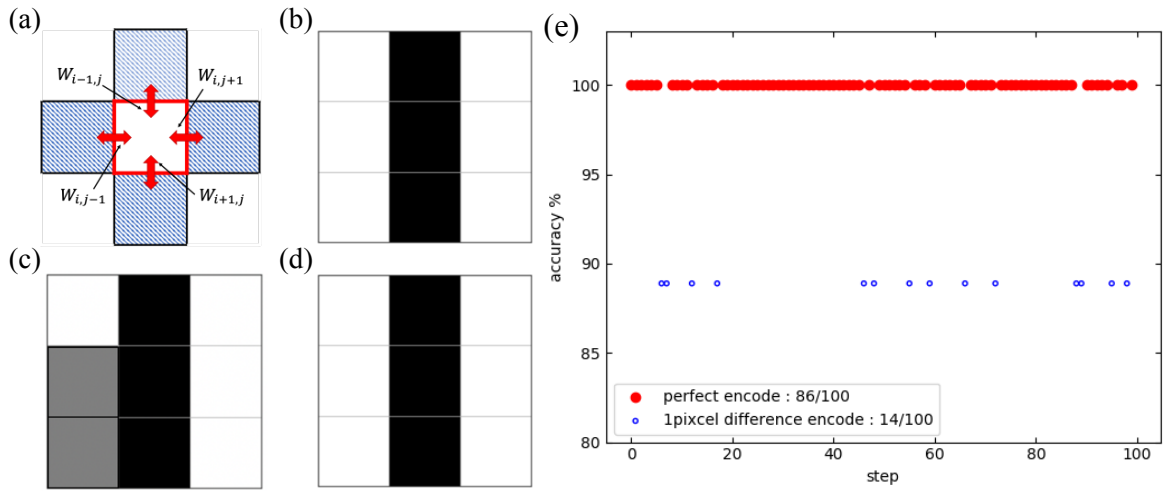

Fig.2 (a) Setting example of coupling coefficient $W(\theta)$ (b) 9pixel original image (c) 9pixel noise mixed image (noise rate 30%) (d) Output image of quantum autoencoder (e) Accuracy of denoising of quantum autoencoder in 9pixel image

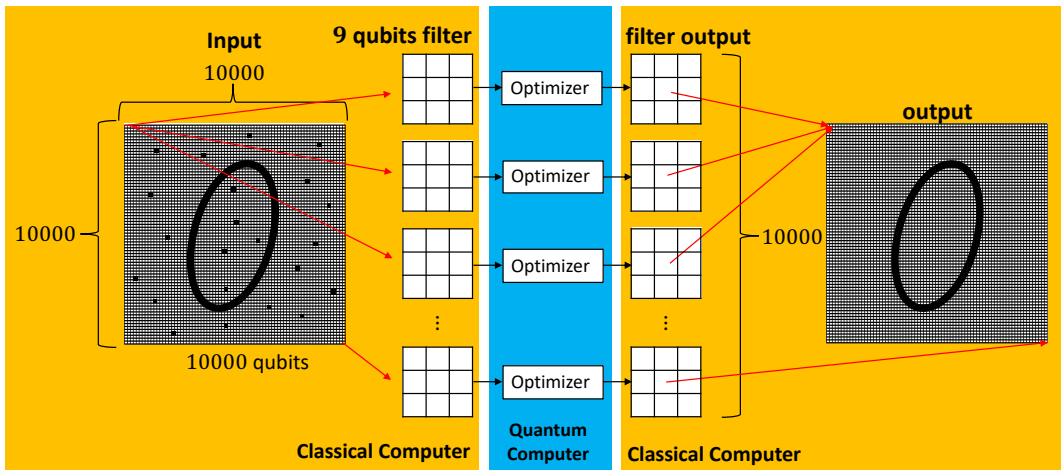

Fig. 3 Conceptual diagram of convolutional quantum autoencoder

At a noise rate of 30%, the original data image could be completely restored at a rate of 86%. Also, it was not possible to completely remove the noise at a rate of 14%. This result is compared with that of quantum autoencoder is performed by a calculation method that applies convolution to quantum computation.

### 2.3  Convolutional quantum autoencoder

As described above, it is difficult to calculate big data in a quantum computer with a limited number of qubits. In order to address this problem, instead of calculating the original image at one time, we consider 9 pixels as a convolution filter as shown in Fig. 3 and propose a method of calculating every one pixel of the original image. Although each pixel interacts only with the nearest neighbors in the upper, lower, left, and right directions as shown in Fig. 2(a), only the central pixel at 9 pixels extracted from the original image is affected by the nearest neighbor same as the original image. Therefore, we calculated 1 pixel each of the original image with a 9pixel filter, and finally made the output image for the original image by collecting the central pixel of each filter. We thought that this method would also function as autoencoder. The quantum circuit of the 9pixel convolution filter in the convolutional quantum autoencoder is the same as in Fig. 1.

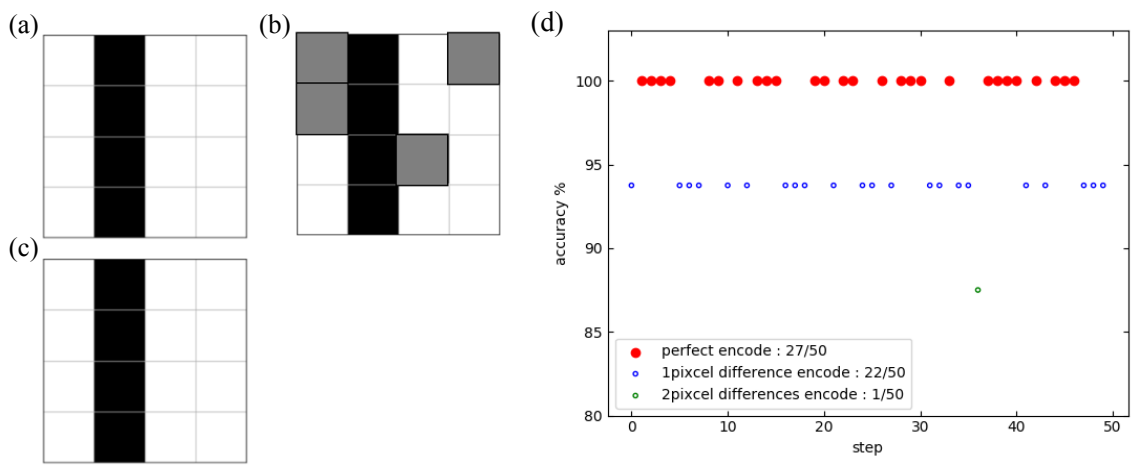

Fig.4 (a) 16pixel original image (b) 16pixel noise mixed image (noise rate 30%) (c) Output image of convolutional quantum autoencoder (d) Accuracy of denoising of convolutional quantum autoencoder in 16pixel image

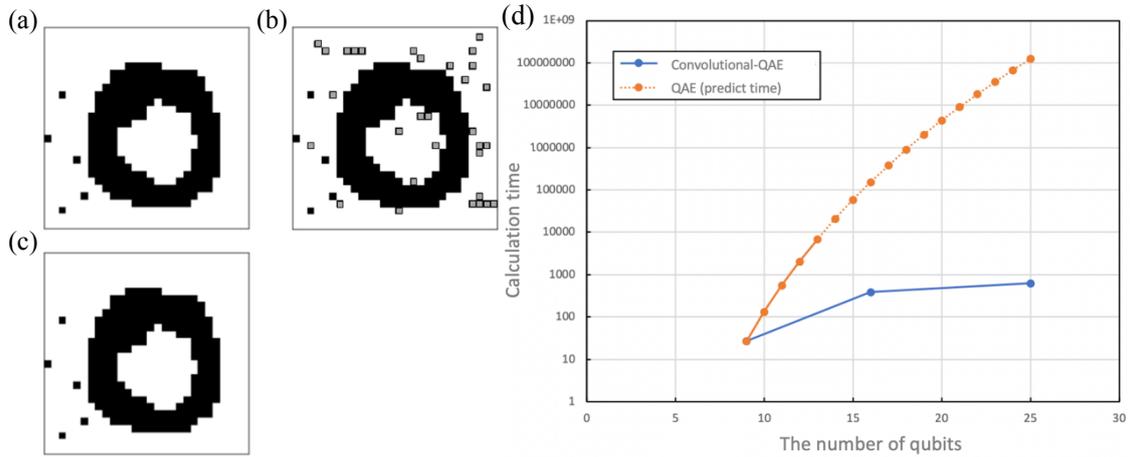

Fig.5 (a) 784pixel original image (b) 784pixel noise mixed image (noise rate 10%) (c) Output image of convolutional quantum autoencoder (d) Computational time comparison of convolutional quantum autoencoder and normal quantum autoencoder in quantum simulator

## 3. Verification of denoising of Image

### 3.1 Verification of denoising of 16pixel Image

We made a 16pixel black-and-white image the original data, and verified the effectiveness of applying convolution to the quantum autoencoder. The original image and the noise mixed image are shown in Fig. 4(a) and 4(b).

The result of having performed noise removal using a quantum adiabatic algorithm in this noise image is shown in Fig. 4(c). In addition, Fig. 4(d) shows the result of testing the noise removal 50 times and verifying the accuracy of the convolutional quantum autoencoder in a 16pixel image.

From Fig. 4(c), it can be seen that the output data image is the same as the original data image of Fig. 4(a). Therefore, it is considered that noise removal can be performed. Further, Fig. 4(d) shows that in the case of a noise rate of 30%, the input data image can be completely restored at a rate of 54%. However, in comparison with Fig. 2(e) which is a normal quantum autoencoder, it can be seen that the accuracy is low. It is assumed that this is because the noise of the original image could be completely removed only when the noise removal was successful in all 16 times of calculation with the 9pixel filter. It was found that the input data image can be restored at a rate of 98% if the noise residual result of 1 pixel is also acceptable.

### 3.2 Verification of denoising of 784pixel (MNIST) Image

Next, we use MNIST's number 0 image as the original data and verify the effectiveness of applying convolution to the quantum autoencoder. The original image and the noise mixed image are shown in Fig. 5(a) and 5(b). The result of performing noise removal using quantum adiabatic algorithm in this noise image is shown in Fig. 5(c).

From Fig. 5(c), it can be seen that the output data image is exactly the same as the original data image of Fig. 5(a). Therefore, it is assumed that even in the MNIST image in which 784 qubits are required in a normal quantum autoencoder, noise removal can be performed with 9 qubits in the convolutional quantum autoencoder.

### 3.3 Verification of calculation time on simulator

Next, on a quantum simulator, we compare the computation time when denoising an image with a convolutional quantum autoencoder and the computation time with a normal quantum autoencoder in Chap. 2.2. In addition, the calculation in a normal quantum autoencoder is to the case of 13 qubits, and the calculation time after it is an estimated value. The results are shown in Fig. 5(d).

Comparing each of them, it was possible to calculate in about 6 minutes in the convolution type method, compared to the estimated value of about 42 hours in the normal method for denoising of a 16pixel image. Also, in the case of denoising a 25pixel image, it can be estimated that it would take about 4 years by the normal method, but it could be calculated in about 10 minutes by the convolution method. From these results, it can be seen that the computational complexity of $O(2^n)$ in a normal quantum autoencoder becomes $O(n)$ in a convolutional quantum autoencoder, and the calculation time can be significantly shortened. Therefore, it is assumed that the convolutional method is effective not only in an actual NISQ device but also in a quantum simulator.

## 4. Verification of dimensionality reduction of input image

The autoencoder has an advantage that it can perform feature extraction by dimension reduction other than image noise removal. In the case of the conventional autoencoder, the output of the intermediate layer can be regarded as data obtained by reducing the dimension of the input data by setting the intermediate layer of the neural network to a dimension lower than that of the input layer.

The quantum autoencoder developed in this research uses the coupling coefficient $W(\theta)$ between spins, which characterizes the input data, and performs dimension reduction by extracting necessary elements from $W(\theta)$. This operation is performed by a classic computer.

As an example, we reduced the dimension of MNIST image to two dimensions with a quantum autoencoder and verified whether it could extract features. First, we prepared coupling coefficients $W(\theta)$ with each MNIST image as input data. In this case, we did not learn $W(\theta)$, but set it directly from the input data. Also, we considered $W(\theta)$ only between the nearest four spins for each spin, but among them, $W(\theta)$ between the right side spin and $W(\theta)$ between the down side spin were extracted. Furthermore, we reduced the dimension by summing each $W(\theta)$ at all spins. In other words, in each pixel of one MNIST image, we have summarized the components of the interaction with the right side pixel and the components of the interaction with the down side pixel. Fig. 6(a) shows a conceptual diagram of dimension reduction in the case of 9 pixels.

The results of the dimensional reduction of the 0 and 1 MNIST images by 30 each using the quantum autoencoder and plotting are shown in Fig. 6(b). We can usually classify 0 and 1 if we reduce the MNIST image of 0 and 1 to 2 dimensions using a conventional autoencoder or Principal Component Analysis (PCA) [2]. From Fig. 6(b), it was found that 0 and 1 can be also classified when we reduce the MNIST image of 0 and 1 to 2 dimensions using a quantum autoencoder.

Further, Fig. 6(c) shows the results of dimensional reduction of the 0 to 9 MNIST images to two dimensions using a quantum autoencoder, and 30 plots of each. In addition, since 0, 1, 3, and 9 MNIST

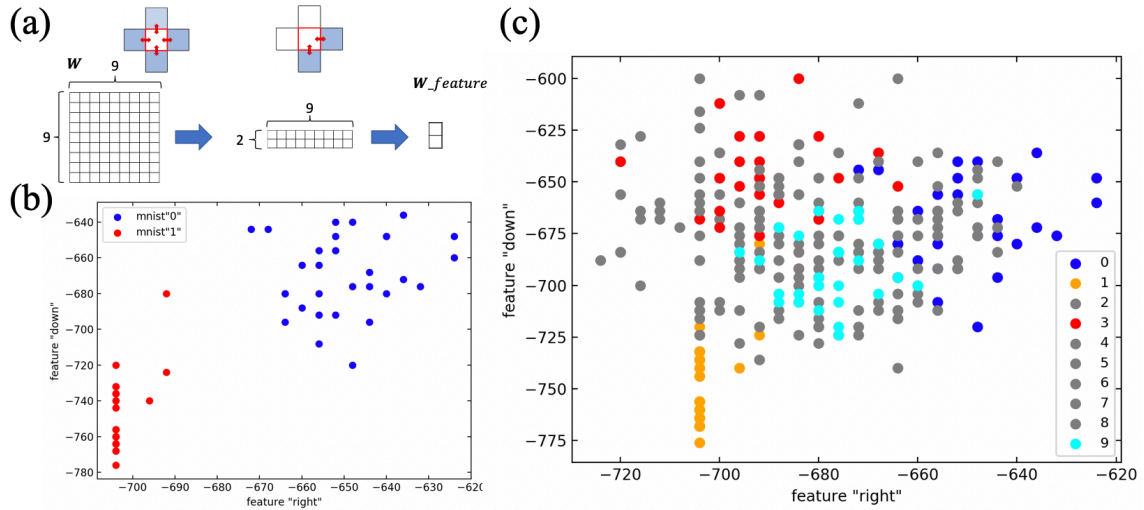

Fig. 6(a) Conceptual diagram of dimensional reduction of quantum autoencoder (b) The result of compressing 0 and 1 MNIST images to 2 dimensions with a quantum autoencoder (c) The result of compressing 0 to 9 MNIST images to 2 dimensions with quantum autoencoder

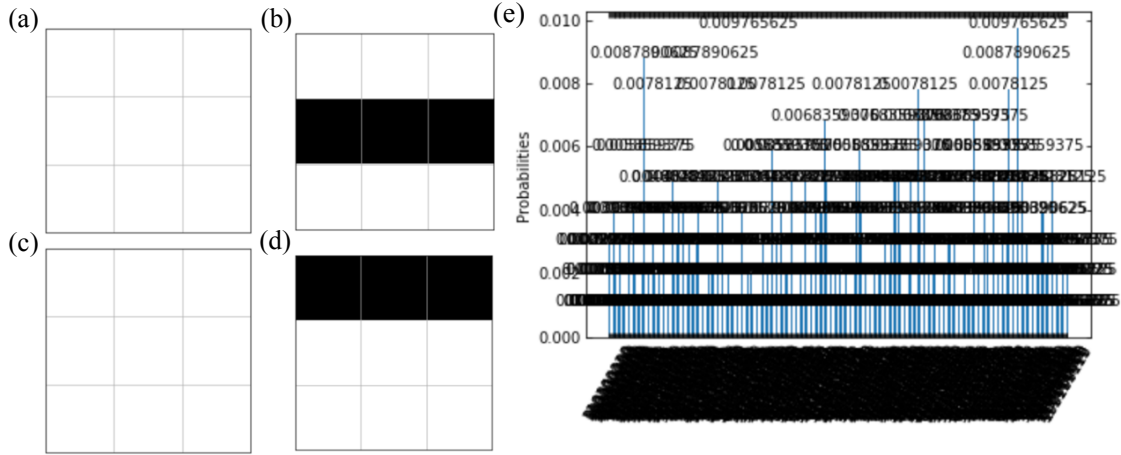

Fig.7 (a) 9pixel original image (b) 9pixel noise mixed image (noise rate 30%) (c) Output image of ReNomQ (d) Output image of IBMQ Melbourne (e) Observation results of qubits in quantum adiabatic algorithm with IBMQ Melbourne images were classified, data points were colored. As shown in Fig. 6(c), even when the image is compressed to two dimensions, differences in what can be classified and those which cannot be classified appear, so it is assumed that the features of each MNIST image can be extracted well.

## 5. Verification of denoising on a real machine

We denoised the 9pixel image using the quantum autoencoder developed in this study using IBM Q Melbourne, which is the hardware of quantum computer. The connection with IBM Q Melbourne was made via Qiskit, a quantum simulator. Furthermore, we also performed the same calculations with the quantum simulator ReNomQ, and compared the results of the NISQ device with the simulator. First, Fig. 7(a) and 7(b) show an original image of 9pixel and a noise mixed image used for verification. Next, the output image of the quantum autoencoder in ReNomQ is shown in Fig. 7(c), and the output image of the quantum autoencoder in IBMQ Melbourne is shown in Fig. 7(d).

From Fig. 7, it can be seen that on the simulator, the output image is equal to the original image, and the noise can be removed, but in the real machine the noise cannot be removed well. Also, Fig. 7(e) shows the observation result of the qubit after the quantum adiabatic algorithm when the quantum autoencoder is calculated by IBMQ Melbourne. It can be seen from Fig. 7(e) that the stable state has not been determined, and there are multiple solution candidates. Therefore, it turned out that IBMQ Melbourne did not calculate correctly. The cause may be that the number of gates in the quantum circuit of the quantum adiabatic algorithm being used is large. Because NISQ devices do not have an error correction function, when the number of gates used increases, errors occur in various places and correct results cannot be obtained.

## 6. Conclusion

In this paper, we proposed a method to apply the convolution method to quantum computing as a computing method of big data in quantum computer with a limited number of qubits. As a result of the experiment, we have found the effectiveness of the convolutional quantum computation, though the accuracy is lower. Furthermore, we have demonstrated that it is possible to learn parameters $W(\theta)$ from input data and to reduce the dimension of input data, as in the conventional autoencoder. Moreover, it could be seen that although the algorithm of this quantum autoencoder gives a good result on a simulator, it does not give a correct result on a real machine. In the future, we would like to work on optimization of quantum circuits as a task so that NISQ devices can also calculate correctly.